\documentstyle[aps,prl,psfig,multicol]{revtex}
\begin{document}
\newenvironment{tab}[1]
{\begin{tabular}{|#1|}\hline}
{\hline\end{tabular}}

\title {BdG equations within lattice Hubbard
model for the description of $\pi$-states in nanoscale S-FF-S junctions
}

\author{N. Stefanakis}
\address{Institut f\"ur Theoretische Physik,
Universit\"at  T\"ubingen,
Auf der Morgenstelle 14, 72076
T\"ubingen Germany}  
\date{\today}
\maketitle

\begin{abstract}
We calculate the persistent currents in
$S-F_{\uparrow}F_{\uparrow[\downarrow]}-S$ 
structures
(S denotes the superconductor in a closed ring 
geometry, F the ferromagnet in a two dimensional geometry, 
and the arrows the alignment of exchange field) as a function
of the applied magnetic flux,
self consistently by solving the Bogoliubov-de Gennes equations within
the two dimensional Hubbard model.
The local current shows a sign change i.e. a $0$ to $\pi$ transition  
in the parallel alignment of the magnetizations 
and it does not change sign in antiparallel alignment. 

\end{abstract}
\section{Introduction}

Presently the study of $\pi$ states is a subject of intensive 
theoretical and experimental investigation. 
The $\pi$ states 
have been predicted theoretically long time ago
in junctions containing magnetic impurities. 
They appear as a sign change of the Josephson 
critical current in SFS structures 
(S denotes the superconductor and F the ferromagnet). 
The crossover from the 
$0$ to $\pi$ state has been observed recently in the critical current 
versus temperature in SFS junctions \cite{ryazanov}, and in the 
critical current versus ferromagnetic layer thickness 
in SIFS junctions \cite{kontos1}. 
More recently the concept of $\pi$ state was used to explain 
the observed shift in the critical versus magnetic field pattern 
by half the flux quantum in the ferromagnetic 
$0-\pi$ SQUID \cite{guichard}, 
and the decaying density of states oscillations 
in $s$-wave superconductor ferromagnet hybrid 
structures \cite{kontos}.
Similar effects have been observed in $d$-wave 
\cite{freamat,freamat1} superconductor ferromagnet hybrid structures.
Theoretical explanation has been given in the framework of the 
quasiclassical theory for $s$-wave \cite{zareyan1} 
and $d$-wave case \cite{zareyan2}. 

Alternatively $\pi$ states appear in Josephson junctions 
involving 
$d$-wave superconductor due to the 
sign change of the superconducting order parameter \cite{kashiwaya}. 
For example the $0$ to $\pi$ transition occurs in the 
critical current versus temperature in Josephson junctions 
with $d$-wave symmetry and depends on the orientation of the lobes 
of the $d$-wave order parameter with respect to the interface. 
The Josephson effect has been studied
in junctions between unconventional 
superconductors across different types of magnetic barriers 
\cite{tanaka}. 

In superconductor - ferromagnet multiterminal hybrid structures 
the proximity effect can be controlled by the 
alignment of magnetizations in 
ferromagnetic electrodes connected to 
superconductor
\cite{melin,jirari,stefan2}
or by the magnetic flux in an Aharonov-Bohm loop 
connected to superconductor \cite{stefan4}.
In two superconductor - ferromanget bilayers separated 
by thin insulating film,
the enhancement of the Josephson critical current
when the orientations of the 
magnetic moments in the ferromagnetic layers are 
antiparallel has been predicted \cite{bergeret}.
Since then several S-FF-S, SF-SF structures have been 
examined 
in the limit of thin F layer and in the diffusive limit 
by solving the Usadel equations \cite{colubov,colubov2}. They found that 
in the case of antiparallel alignment of magnetizations in the 
ferromagnetic electrodes the critical current enhances with 
the exchange field, while in the parallel alignment 
the junction exhibits a transition to the $\pi$ state \cite{colubov}.
Also non-sinusoidal current phase relation in S-F-S, SF-c-SF 
junctions has been predicted \cite{colubov2}. 
The same problem has been treated in the ballistic regime \cite{blanter}.

In this 
paper 
our goal is to explore several new aspects
related to the control of the Josephson effect in
nanostructures. 
We study
two dimensional $F_{\uparrow} F_{\uparrow[\downarrow]}$
embedded in
a superconducting wire.
The method is based on exact diagonalizations
of the Bogoliubov-de Gennes equations associated to the mean field
solution 
of an extended Hubbard model.
The basic quantity which we calculate is
the local current
as a function of 
several relevant parameters:
the distance from the surface, the magnetic field that 
penetrates the superconducting loop and exchange field.

We find that the local
current is periodic function
of the flux and is controlled by the 
orientation of magnetization in the ferromagnetic electrodes.
The local current shows a sign change i.e. a $0$ to $\pi$ transition
in the parallel alignment of the magnetizations
and it does not change sign in antiparallel alignment.
Our predictions from the simulations of this model are of interest in
view of future experiments on nanoscale Josephson junctions.

The article is organized as follows. In Sec. II we
develop the model and discuss the formalism. In Sec. III  
we discuss metallic or ferromagnetic structures embedded inside
a superconducting loop.
Finally summary and discussions are presented in the last section.

\section{BdG equations
within the Hubbard model}

The Hamiltonian for the Hubbard lattice model
is
\begin{eqnarray}
H & = & -t\sum_{<i,j>\sigma}c_{i\sigma}^{\dagger}c_{j\sigma}
+\mu \sum_{i\sigma} n_{i\sigma}
+\sum_{i\sigma} h_{i\sigma}n_{i\sigma} \nonumber \\
  & + & V_0\sum_{i} n_{i\uparrow} n_{i\downarrow}
,~~~\label{bdgH}
\end{eqnarray}
where $i,j$ are sites indices and the angle brackets indicate that the
hopping is only to nearest neighbors,
$n_{i\sigma}=c_{i\sigma}^{\dagger}c_{i\sigma}$ is the electron number
operator in site $i$, $\mu$ is the chemical potential which is 
set to zero.
$h_{i\sigma}=-h\sigma_z$, is the exchange field
in the ferromagnetic region
and $\sigma_z=\pm 1$ is the eigenvalue of the
$z$ component of the Pauli matrix.
$V_0$ is
the on site interaction strength which gives rise to
superconductivity.
Within the mean field approximation Eq. (\ref{bdgH}) is reduced  to
the Bogoliubov deGennes equations \cite{gennes}:
\begin{equation}
\left(
\begin{array}{ll}
  \hat{\xi} & \hat{\Delta} \\
  \hat{\Delta}^{\ast} & -\hat{\xi}
\end{array}
\right)
\left(
\begin{array}{ll}
  u_{n \uparrow}(r_i) \\
  v_{n \downarrow}(r_i)
\end{array}
\right)
=\epsilon_{n\gamma_1}
\left(
\begin{array}{ll}
  u_{n \uparrow}(r_i) \\
  v_{n \downarrow}(r_i)
\end{array}
\right)
,~~~\label{bdgbdg1}
\end{equation}

\begin{equation}
\left(
\begin{array}{ll}
  \hat{\xi} & \hat{\Delta} \\
  \hat{\Delta}^{\ast} & -\hat{\xi}
\end{array}
\right)
\left(
\begin{array}{ll}
  u_{n \downarrow}(r_i) \\
  v_{n \uparrow}(r_i)
\end{array}
\right)
=\epsilon_{n\gamma_2}
\left(
\begin{array}{ll}
  u_{n \downarrow}(r_i) \\
  v_{n \uparrow}(r_i)
\end{array}
\right)
,~~~\label{bdgbdg2}
\end{equation}

such that
\begin{equation}
\hat{\xi}u_{n\sigma}(r_i)=-t\sum_{\hat{\delta}}
u_{n\sigma}(r_i+\hat{\delta})+\mu u_{n\sigma}(r_i)+
h_i\sigma_z u_{n\sigma}(r_i),~~~\label{bdgxi}
\end{equation}

\begin{equation}
\hat{\Delta}u_{n\sigma}(r_i)=\Delta_0(r_i)u_{n\sigma}(r_i),
~~~\label{bdgdelta}
\end{equation}
where the pair potential is defined by
\begin{equation}
\Delta_0(r_i)\equiv
V_0<c_{\uparrow}(r_i)c_{\downarrow}(r_i)>.~~~\label{bdgdelta0}
\end{equation}
Equations
(\ref{bdgbdg1},\ref{bdgbdg2})
are subject to the self consistency requirement
\begin{equation}
\Delta_0(r_i)  =  V_0(r_i)F(r_i)=
\frac{V_0(r_i)}{2}
\sum_{n} \left[
u_{n\uparrow}(r_i)v_{n\downarrow}^{\ast}(r_i)\tanh(\beta
\epsilon_{n\gamma_1}/2))+
u_{n\downarrow}(r_i)v_{n\uparrow}^{\ast}(r_i)\tanh(\beta
\epsilon_{n\gamma_2}/2) \right]
,~~~\label{bdgselfD0}
\end{equation}

$F(r_i)$ is the pair amplitude.
We solve the above equations self consistently.
The numerical procedure has been described elsewhere
\cite{jirari,stefan2,stefan,tanuma,zhu1}.

The local current between sites $i,j$ is given by
\begin{equation}
I_{ij}=\frac{i e}{\hbar}(t<c_i^{\dagger}c_j>-t^{*}<c_j^{\dagger}c_i>)
\end{equation}.

\section{Results}

\begin{figure}
\begin{center}
\leavevmode
\psfig{figure=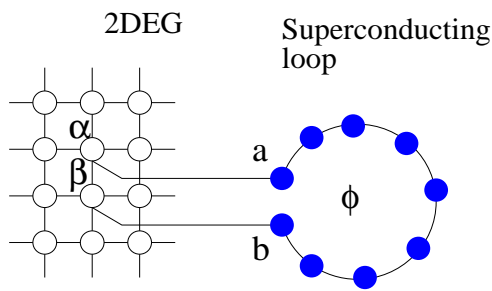,width=6.5cm,angle=0}
\end{center}
\caption{The SNS junction of the superconducting wire with 
the 2DEG. The loop contains 20 sites while the 2DEG is of
$8 \times 8$ sites. The labeling of several sites close to the 
interface
is shown.
}
\label{mesoloop.fig}
\end{figure}

We study the quasiparticle 
properties of a two dimensional plane to which a 
superconducting wire is attached at two points and a magnetic flux 
is applied through the wire. The magnetic flux
creates a phase difference between sites $\alpha$ and $\beta$
(see Fig. \ref{mesoloop.fig}), which is proportional to the 
applied flux. Alternatively a current through the 
wire can create this phase difference.   
We demonstrate in this section that the magnetic flux through
the wire
as seen in Fig. \ref{mesoloop.fig}
can be used to control the proximity effect in
this hybrid structure.
The magnetic flux through the loop is modeled as a factor
$e^{i f}$ where $f=2\pi \Phi/ \Phi_0 =2\pi \phi$
in the hopping integral. $\Phi_0=\frac{h}{e}$ is the 
normal flux quantum.
In the calculation we used a small cluster
of $8\times 8$ sites to model the 2DEG and also
open boundary conditions, while a chain of $20$ sites models 
the superconducting wire.
The hybrid normal-superconducting ring has been discussed in Refs. 
\cite{buttiker,cayssol}.
They found that the persistent currents depend on the length 
of the N segment compared to the S. For large length of the 
superconducting region the quasiparticle wave function is 
localized in the normal metal region and quasiparticles are 
entering the superconductor via Andreev reflection. In this case 
the persistent currents have period half the
flux quantum. In the opposite limit where the 
length of the 
superconducting region is small compared to the coherence length
the quasiparticle wave function
becomes extended and the persistent currents have period 
equal to the
flux quantum. In our case the quasiparticle properties 
are more close to the former case.
We checked the case where the wire and 
the reservoir are both normal metals. In that case 
we found that the persistent currents have period $\Phi_0$. 
Also the phase of the persistent currents is zero in a loop 
with even number of carriers and 
equal to $\pi$ in a loop with odd number of carriers. 
However in our case we fix the loop size and study the 
response of the system to the magnetic field and exchange field.
We distinguish the following cases depending on the quality of the 
barrier

\subsection{S-N-S}

\begin{figure}
\begin{center}
\leavevmode
\psfig{figure=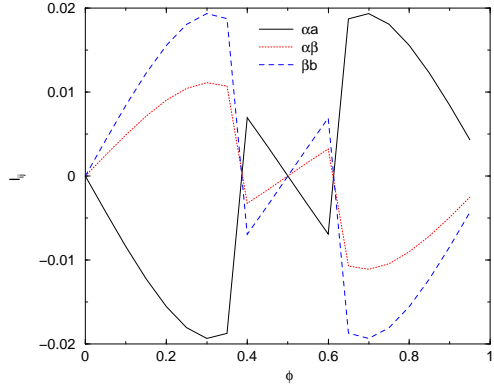,width=6.5cm,angle=0}
\end{center}
\caption{
The local current $I_{ij}$ between sites $i,j$ of the SNS junction,
where $i j=$ $\alpha a$, $\alpha \beta$, $\beta b$ 
as a function of the magnetic flux. The exchange field is equal to $h=0$
and the value of the superconducting gap is $V_0=-3.5$.
}
\label{n.fig}
\end{figure}

We discuss first the case where a superconducting wire is connected
to a 2DEG as seen in Fig. \ref{mesoloop.fig}. 
We demonstrate that the flux through the loop modulates in a 
periodic way the quasiparticle properties of the film, so that 
the proximity effect is controlled by the applied flux. 
The local current-magnetic flux relation is seen 
in Fig. \ref{n.fig} between sites
$\alpha a$, $\alpha \beta$ and $\beta b$.
It shows the characteristic sinusoidal form,
with period almost $\Phi_0/2$.

\subsection{S-F$_{\uparrow}$F$_{\uparrow}$-S}

\begin{figure}
\begin{center}
\leavevmode
\psfig{figure=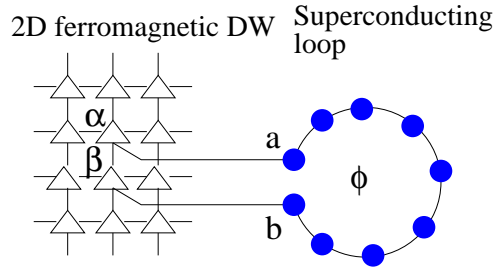,width=6.5cm,angle=0}
\end{center}
\caption{The S-F$_{\uparrow}$F$_{\uparrow}$-S 
junction of the superconducting wire with 
the ferromagnetic 2DEG. At the connections sites $\alpha,\beta$ the 
ferromagnetic atoms have the same orientation of magnetizations. 
The loop contains 20 sites while the 2DEG is of
$8 \times 8$ sites.  
}
\label{mesoloopF.fig}
\end{figure}

\begin{figure}
\begin{center}
\leavevmode
\psfig{figure=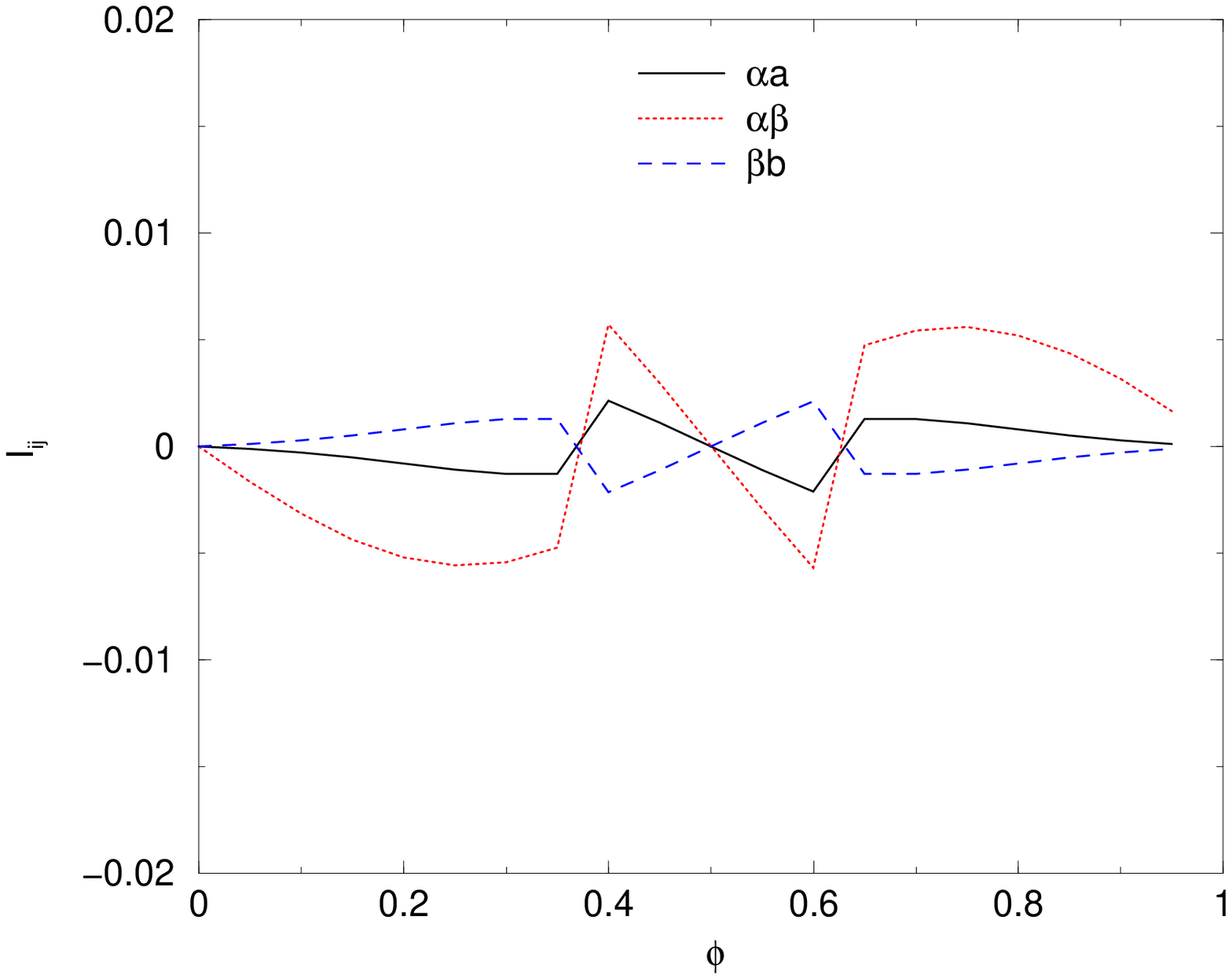,width=6.5cm,angle=0}
\end{center}
\caption{
The local current $I_{ij}$ between sites $i,j$ of the 
S-F$_{\uparrow}$F$_{\uparrow}$-S junction,
where $i j=$ $\alpha a$, $\alpha \beta$,  $\beta b$ 
as a function of the magnetic flux. The exchange field is equal to $h=3$
and the value of the superconducting gap is $V_0=-3.5$.
}
\label{f.fig}
\end{figure}

We discuss now the case where a superconducting wire is connected
to a ferromagnetic film as seen in Fig. \ref{mesoloopF.fig}. 
The objective is to investigate the effect of the exchange field in the 
current. The 
ferromagnetic atoms at sites $\alpha$ and $\beta$ where the superconducting 
wire is connected to the film have the same orientation 
of magnetizations.
Due to proximity effect
the pair amplitude shows decaying oscillations with alternating 
positive and negative signs inside the ferromagnet
away from the connection points $\alpha,\beta$.
The local current 
oscillates with $\phi$
as seen in Fig. \ref{f.fig}. 
The Josephson effect occurs due to
transfer of Cooper pairs across the interface via
Andreev reflection. In the ferromagnetic site with spin up orientation
of magnetization $\alpha$ and in the
ballistic limit the phase difference between the electron and
Andreev reflected hole depends on the exchange field
i.e. $\delta \phi=2hx/\hbar v_F$.
In the other ferromagnetic site with the same orientation
of magnetization $\beta$ the phase shift has the same sign.
So in total in the parallel alignment of magnetizations
there is a phase shift which can be equal to $\pi$, 
and the critical current is
expected to change sign compared to the SNS case.
Indeed $I_{\alpha\beta}$ has opposite sign in relation to 
the normal metal case as seen in Fig. \ref{f.fig}.

\subsection{S-F$_{\uparrow}$F$_{\downarrow}$-S}

\begin{figure}
\begin{center}
\leavevmode
\psfig{figure=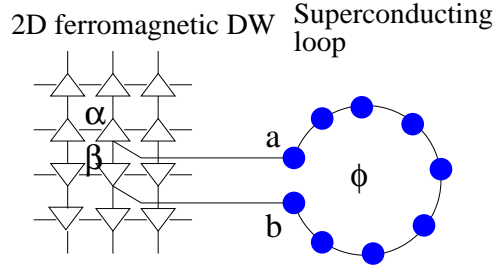,width=6.5cm,angle=0}
\end{center}
\caption{The S-F$_{\uparrow}$F$_{\downarrow}$-S 
junction of the superconducting wire with 
the ferromagnet with antiparallel orientations of the 
magnetizations. The loop contains 20 sites while the 2DEG is of
$8 \times 8$ sites.  
}
\label{mesoloopAF.fig}
\end{figure}

\begin{figure}
\begin{center}
\leavevmode
\psfig{figure=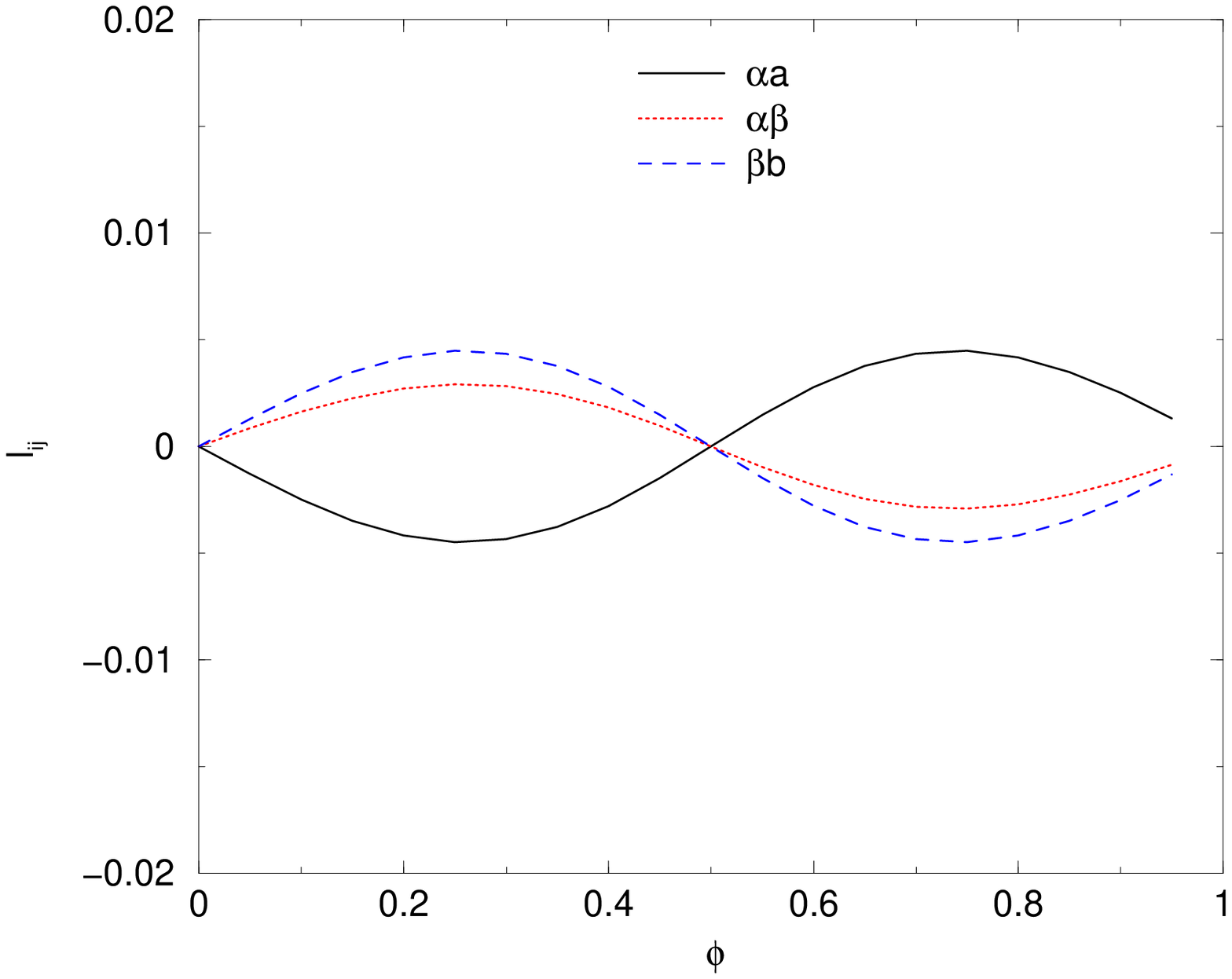,width=5.5cm,angle=0}
\end{center}
\caption{
The local current $I_{ij}$ between sites $i,j$ of the 
S-F$_{\uparrow}$F$_{\downarrow}$-S junction, 
where $i j=$ $\alpha a$, $\alpha \beta$, $\beta b$ 
as a function of the magnetic flux. The exchange field is equal to $h=3$
and the value of the superconducting gap is $V_0=-3.5$.
}
\label{af.fig}
\end{figure}

We discuss now the case where a superconducting wire is connected
to a two dimensional 
ferromagnetic domain wall as seen in Fig. \ref{mesoloopAF.fig}. 
We demonstrate that the alignment of the magnetization provides 
additional control parameter for the proximity effect.
The
ferromagnetic atoms at sites $\alpha$ and $\beta$ where the superconducting 
wire is attached have the opposite orientation
of magnetizations.
The local current 
oscillates with $\phi$
as seen in Fig. \ref{af.fig}. 
However 
differently to the parallel alignment case, 
the local current $I_{\alpha\beta}$ 
does not change sign with the increase of the exchange field.
These results indicate that the junction in the 
antiparallel alignment of magnetizations strongly resembles the 
SNS junction. 
We emphasize here that the ferromagnetic layers are small 
(one atomic layer) so there is no spatial variation of the 
pairing amplitude inside the ferromagnet. Therefore the $0$ to 
$\pi$ transition is explained in terms of phase discontinuities 
at the interface. 
For the antiparallel alignment the phase shift across the junctions is 
zero. The Josephson effect can be viewed as a 
transfer of Cooper pairs across the interface via 
Andreev reflection. In the ferromagnetic site with spin up orientation 
of magnetization $\alpha$ and in the 
ballistic limit the phase difference between the electron and 
Andreev reflected hole depends on the exchange field 
i.e. $\delta \phi=2hx/\hbar v_F$. 
In the other ferromagnetic site with spin down orientation
of magnetization $\beta$ the phase shift is the opposite. 
So in total in the antiparallel alignment of magnetizations 
there is not a phase shift, and the local current is 
expected to be similar to the SNS case. 
For the parallel alignment 
the phase shifts are added and this provides the possibility for a $\pi$ state. 
This $\pi$ phase shift in the phase of the order parameter 
is responsible for the sign change of the current. 
Similar analytical results have been obtained using the 
Usadel equations \cite{colubov,colubov2}. 
However in that case several approximations were used like 
constant order parameter in the superconducting electrodes
and strongly coupled SF interfaces.

\section{conclusions}
We calculated the local current for
normal and ferromagnetic planes with superconducting mirrors,
in the presence of an 
external magnetic field
within the extended Hubbard 
model, self consistently. 
The local current, is periodic function
of the magnetic flux that is applied through the superconducting wire. 
The local current versus exchange field is further controlled by the 
alignment of the magnetizations in the ferromagnetic electrodes. 
Specifically it
shows a sign change in the parallel alignment of the 
magnetizations 
but it does not change sign in the antiparallel alignment. 

Our results are influenced from the details of the 
lattice structure and quantitative differences are found 
compared to the results of quasiclassical theory.
The results can be generalized to extended contacts. We expect only 
quantitative differences. 
In the present paper we used superconducting rings where the applied flux 
creates a phase difference in the interruption points. 
However the results can be generalized 
to junctions instead of rings, where the current
creates a phase difference between the 
two superconductors. 

\bibliographystyle{prsty}

\begin{thebibliography}{99}

\bibitem{ryazanov} V.V. Ryazanov, V.A. Oboznov, A.Yu. Rusanov, 
A.V. Veretennikov, A.A. Golubov, and J. Aarts, 
Phys. Rev. Lett. {\bf 86}, 2427 (2001).

\bibitem{kontos1} T. Kontos, M. Aprili, J. Lesueur,  
F. Gen\^et, B. Stefanidis, and R. Boursier, 
Phys. Rev. Lett. {\bf 89}, 137007 (2002).

\bibitem{guichard} W. Guichard, M. Aprili, O. Bourgeois, T. Kontos, 
J. Lesueur, and P. Gandit,
Phys. Rev. Lett. {\bf 90}, 167001 (2003).

\bibitem{kontos} T. Kontos, M. Aprili, J. Lesueur, and X. Grison,
Phys. Rev. Lett. {\bf 86}, 304 (2001).

\bibitem{freamat} M. Freamat and K.-W. Ng,
cond-mat/0301081.

\bibitem{freamat1} M. Freamat and K.-W. Ng,
cond-mat/0305446.

\bibitem{zareyan1} M. Zareyan, W. Belzig, and Yu.V. Nazarov,
Phys. Rev. Lett. {\bf 86}, 308 (2001).

\bibitem{zareyan2} Z. Faraii and M. Zareyan,
cond-mat/0304336.

\bibitem{kashiwaya} S. Kashiwaya and Y. Tanaka,
Rep. Prog. Phys. {\bf 63}, 1641 (2000).

\bibitem{tanaka} Y. Tanaka and S. Kashiwaya,
J. Phys. Soc. Jpn. {\bf 68}, 3485 (1999).

\bibitem{melin} R. M\'elin and D. Feinberg,
Eur. Phys. J. B {\bf 26}, 101 (2002).

\bibitem{jirari} H. Jirari, R. M\'elin and N. Stefanakis,
Eur. Phys. J. B {\bf 31}, 125 (2003).

\bibitem{stefan2} N. Stefanakis and R. M\'elin,
J. Phys. Condens. Matter {\bf 15}, 3401 (2003).

\bibitem{stefan4} N. Stefanakis,
cond-mat/0306348.

\bibitem{bergeret} F.S. Bergeret, A.F. Volkov, and K.B. Efetov,
Phys. Rev. Lett. {\bf 86}, 4096 (2001).

\bibitem{colubov} A.A. Colubov, M.Yu. Kupriyanov, and Ya.V. Fominov,
JETP Lett. {\bf 75}, 190 (2002).

\bibitem{colubov2} A.A. Colubov, M.Yu. Kupriyanov, and Ya.V. Fominov,
JETP Lett. {\bf 75}, 588 (2002).

\bibitem{blanter} Ya.M. Blanter and F.W.J. Hekking,
cond-mat/0306706.

\bibitem{gennes} P.G. de Gennes, {\em Superconductivity of Metals and Alloys} 
(Benjamin, New York, 1966).

\bibitem{stefan} N. Stefanakis, Phys. Rev. B
{\bf 66}, 024514 (2002).

\bibitem{tanuma} Y. Tanuma, Y. Tanaka, M. Yamashiro, and S. Kashiwaya,
Phys. Rev. B {\bf 57}, 7997 (1998).

\bibitem{zhu1} J.-X. Zhu and C.S. Ting,
Phys. Rev. B {\bf 61}, 1456 (1999).

\bibitem{buttiker} M. B\"uttiker and T.M. Klapwick,
Phys. Rev. B {\bf 33}, 5114 (1986).

\bibitem{cayssol} J. Cayssol, T. Kontos, and G. Montambaux,
Phys. Rev. B {\bf 67}, 184508 (2003).

\end{thebibliography}


\end{document}